\documentclass[cits]{PoS}
\usepackage{amsmath}
\usepackage{amsfonts}
\usepackage{graphicx}
\usepackage{slashed}
\usepackage{dcolumn}
\usepackage{bm}
\newcommand{\laem}{\stackrel{<}{\sim}}
\newcommand{\gaem}{\stackrel{>}{\sim}}

\title{Technicolor and Lattice Gauge Theory\thanks{Portions of this manuscript have previously appeared in \protect\cite{Chivukula:2010tn}, and a more detailed discussion can be found there.}}

\ShortTitle{Technicolor and Lattice Gauge Theory}

\author{\speaker{R. Sekhar Chivukula} and Elizabeth H. Simmons\\
        Department of Physics and Astronomy, Michigan State University, East Lansing, MI 48824 USA\\
        E-mail: \email{sekhar@msu.edu and esimmons@msu.edu}}

\abstract{Technicolor and other theories of dynamical electroweak symmetry breaking invoke
chiral symmetry breaking triggered by strong gauge-dynamics, analogous to that found in {\it QCD},  
to explain the observed $W$, $Z$, and fermion masses. In this talk we 
describe why a realistic theory of dynamical electroweak symmetry
breaking must, relative to {\it QCD}, produce an enhanced fermion condensate. We quantify
the degree to which the technicolor condensate must be enhanced in order to yield the observed quark masses, 
and still be consistent with phenomenological constraints on flavor-changing neutral-currents.
Lattice studies of  technicolor and related theories provide the only way to demonstrate that such enhancements
are possible and, hopefully, to discover viable candidate models. We comment briefly on the current
status of non-perturbative investigations of dynamical electroweak symmetry breaking, and provide
a "wish-list" of phenomenologically-relevant properties that are important to calculate in these theories.}

\FullConference{The XXVIII International Symposium on Lattice Field Theory, Lattice2010\\
		June 14-19, 2010\\
		Villasimius, Italy}

\begin{document}

\section{Technicolor and Extended Technicolor}

The earliest models \cite{Weinberg:1979bn,Susskind:1978ms,Hill:2002ap} of dynamical
electroweak symmetry breaking include a new
asymptotically free
non-abelian gauge theory (``technicolor'') and additional massless fermions
(``technifermions'' transforming under a vectorial
representation of the gauge group) which feel this new force. The global chiral symmetry of
the fermions is spontaneously broken by the formation of a technifermion
condensate, just as the approximate chiral SU$(2) \times SU$(2) symmetry in
QCD is broken down to SU(2) isospin by the formation of a quark
condensate. If the quantum numbers of the technifermions are chosen correctly
({\it e.g.} by choosing technifermions in the fundamental representation of
an SU$(N)$ technicolor gauge group, with the left-handed technifermions being
weak doublets and the right-handed ones weak singlets), this condensate can
break the electroweak interactions down to electromagnetism.

While technicolor chiral symmetry breaking can give mass to the $W$
and $Z$ particles, additional interactions must be introduced to
produce the masses of the standard model fermions. The most thoroughly
studied mechanism for this invokes ``extended technicolor'' (ETC)
gauge 
interactions \cite{Dimopoulos:1979es,Eichten:1979ah}.
In ETC,
technicolor and flavor are embedded into a larger gauge group,
which is broken at a sequence of mass scales down to the
residual, exact technicolor gauge symmetry.  The massive gauge bosons associated with this breaking
mediate transitions between quarks/leptons and technifermions, giving
rise to the couplings necessary to produce fermion masses.  

As noted by Eichten and Lane \cite{Eichten:1979ah},
however, the additional interactions introduced to generate ordinary fermion masses cannot
be flavor-universal, and would therefore also generically give rise to flavor-changing
neutral-current (FCNC) processes. In particular they showed that, absent any ``GIM-like" mechanism 
\cite{Chivukula:1987py,D'Ambrosio:2002ex,Appelquist:2004ai} for suppressing 
flavor-changing neutral currents, the ETC scale associated with strange-quark mass generation
must be larger than of order $10^3$ TeV in order to avoid unacceptably large ($CP$-conserving) contributions to neutral
$K$-meson mixing. To obtain quark masses that are large enough therefore requires an
enhancement of the technifermion condensate over that expected naively
by scaling from QCD. Such an enhancement can occur in ``walking" technicolor theories \cite{Holdom:1981rm,Holdom:1984sk,Yamawaki:1985zg,Appelquist:1986an,Appelquist:1986tr,Appelquist:1987fc}
in which the gauge coupling runs very slowly,\footnote{For some examples of proposed models of walking technicolor, see \protect\cite{Appelquist:1997fp} and \protect\cite{Sannino:2009za} and references therein.} or in ``strong-ETC" theories \cite{Appelquist:1988as,Takeuchi:1989qa,Miransky:1988gk,Fukano:2010yv} in which
the ETC interactions themselves are strong enough to help drive technifermion chiral symmetry breaking.\footnote{It is also notable that walking technicolor
and strong-ETC theories are quite different from QCD, and may be far less constrained by precision
electroweak measurements \protect\cite{Peskin:1990zt,Peskin:1991sw,Hill:2002ap}.}

\section{Constraints on $\Lambda_{ETC}$ from neutral meson mixing}\label{sec:FCNC}

At low energies, the flavor-changing four-fermion interactions induced by ETC boson exchange alter the predicted rate of neutral meson mixing.  Ref. \cite{Bona:2007vi} has derived constraints on general $\Delta F = 2$ four-fermion operators that affect neutral Kaon, D-meson, and B-meson mixing, including the effects of running from the new physics scale down to the meson scale and interpolating between quark and meson degrees of freedom.   Their limits on the coefficients ($C^1_j$) of the FCNC operators involving LH current-current interactions:
\begin{eqnarray}
C^1_K (\bar{s}_L \gamma^\mu d_L) (\bar{s}_L \gamma_\mu d_L) \label{eq:cdeffs}\\
C^1_D\, (\bar{c}_L \gamma^\mu u_L) (\bar{c}_L \gamma_\mu u_L) \\
C^1_{B_d} (\bar{b}_L \gamma^\mu d_L) (\bar{b}_L \gamma_\mu d_L) \\
\! C^1_{B_s} (\bar{b}_L \gamma^\mu s_L) (\bar{b}_L \gamma_\mu s_L) \,,\!
\end{eqnarray}
are listed in the left column of Table \ref{tab:one}. 
In the case of an ETC model with arbitrary flavor structure and no assumed ETC contribution to CP-violation,  one has\footnote{Here we assume there is no flavor symmetry suppressing tree-level flavor-changing neutral
currents \protect\cite{Chivukula:2010tn,Bona:2007vi}.}  $C^1_i = \Lambda_{ETC}^{-2}$ and the limits on the $\Lambda_{ETC}$ from \cite{Bona:2007vi} are as shown in the right-hand column of Table \ref{tab:one}.  The lower bound on $\Lambda_{ETC}$ from $D$-meson mixing is now the strongest, with that from Kaon mixing a close second and those from $B$-meson mixing far weaker.  Since the charm quark is so much heavier than the strange quark, requiring an ETC model to produce $m_c$ from interactions at a scale of over $1000$ TeV is a significantly stronger constraint on model-building than the requirement of producing $m_s$ at that scale.\footnote{Note that if one, instead, assumes that ETC contributes to CP-violation in the Kaon system, then the relevant bound on $\Lambda_{ETC}$ comes from the imaginary part of $C^1_K$  and is a factor of ten more severe (see last row of Table \ref{tab:one}).}

\begin{table}[h]
\caption{Limits from the UTFit Collaboration \cite{Bona:2007vi} on coefficients of left-handed four-fermion operators contributing to neutral meson mixing (left column) and the implied lower bound on the ETC scale (right column).  The bounds in the first four rows apply when one assumes ETC does not contribute to CP violation; the bound in the last row applies if one assumes that ETC does contribute to CP violation in the Kaon system. }
\begin{center}
\tiny
\begin{tabular}{|c|c|} \hline
$\phantom{\dfrac{\strut}{\strut}} $\ \ Bound on operator coefficient\  (GeV$^{-2}$)\ \ &  \ \ Implied lower limit on ETC scale \ ($10^3$ TeV)\ \  \\  \hline\hline
$\phantom{\dfrac{\strut}{\strut}} - 9.6 \times 10^{-13} < \Re(C^1_K) < 9.6 \times 10^{-13}$  & $1.0$ \\  \hline 
$\phantom{\dfrac{\strut}{\strut}}  \vert C^1_D \vert  < 7.2 \times 10^{-13}$ & $1.5$ \\  \hline
$\phantom{\dfrac{\strut}{\strut}}\vert C^1_{B_d} \vert  < 2.3 \times 10^{-11}$ & $0.21$ \\ \hline
$\phantom{\dfrac{\strut}{\strut}}\vert C^1_{B_s} \vert  < 1.1 \times 10^{-9}$ & $0.03$\\  \hline \hline
$\phantom{\dfrac{\strut}{\strut}}- 4.4 \times 10^{-15} < \Im(C^1_K) < 2.8 \times 10^{-15}$ & $10$ \\  \hline
\end{tabular}
\end{center}
\label{tab:one}
\end{table}


\section{Condensate Enhancement and $\gamma_m$}\label{sec:gammam}

In studying how ETC theories produce quark masses, the primary operator of interest has the form\footnote{In an ETC gauge theory, we would expect
$1/\Lambda^2_{ETC} \equiv g^2_{ETC}/M^2_{ETC}$ where $g_{ETC}$ and $M_{ETC}$ are
the appropriate extended technicolor coupling and gauge-boson mass, respectively. At energies below
$M_{ETC}$, these parameters always appear (to leading order in the ETC interactions) in this
ratio -- and therefore, we use $\Lambda_{ETC}$ for simplicity.}
\begin{equation}
\frac{(\bar{Q}^a_L \gamma^\mu q^j_L)(u^i_R \gamma_\mu U^a_R)}{\Lambda^2_{ETC}}~,
\end{equation}
where the $Q^a_L$ and $U^a_R$ are technifermions ($a$ is a technicolor index), and the
$q^j_L$ and $u^i_R$ are left-handed quark doublet and right-handed up-quark gauge-eigenstate
fields ($i$ and $j$ are family indices). This operator will give rise, after technifermion chiral symmetry breaking at
the weak scale, to a fermion mass term of order 
\begin{equation}
{\cal M}_{ij} = \frac{\langle \bar{U}_L U_R \rangle_{\Lambda_{ETC}}}{\Lambda^2_{ETC}}~.
\label{eq:i}
\end{equation}
Here it is important to note that the technifermion condensate, $\langle \bar{U}_L U_R \rangle_{\Lambda_{ETC}}$
is renormalized at the ETC scale \cite{Holdom:1981rm,Holdom:1984sk,Yamawaki:1985zg,Appelquist:1986an,Appelquist:1986tr,Appelquist:1987fc}. It is related to the condensate at the technicolor (electroweak symmetry
breaking) scale by
\begin{equation}
\langle \bar{U}_L U_R \rangle_{\Lambda_{ETC}} = 
\exp\left(\int_{\Lambda_{TC}}^{\Lambda_{ETC}}\gamma_m(\alpha_{TC}(\mu))\frac{d\mu}{\mu}\right)
\langle \bar{U}_L U_R \rangle_{\Lambda_{TC}}~,
\label{eq:ii}
\end{equation}
where $\gamma_m(\alpha_{TC}(\mu))$ is the anomalous dimension of the technifermion mass operator.\footnote{For a discussion of the potential scheme-dependence of $\gamma_m$, see \protect\cite{Chivukula:2010tn}.}
Using an estimate of the technifermion condensate, and a calculation of the anomalous dimension of
the mass operator, we may estimate the size of quark mass which can arise in a technicolor theory for a
given ETC scale.

In a theory of walking technicolor \cite{Holdom:1981rm,Holdom:1984sk,Yamawaki:1985zg,Appelquist:1986an,Appelquist:1986tr,Appelquist:1987fc}, the gauge coupling runs very slowly just above the technicolor scale
$\Lambda_{TC}$. The largest enhancement occurs in the limit of ``extreme walking"
in which the technicolor coupling, and hence the anomalous dimension $\gamma_m$, remains
approximately constant from the technicolor scale, $\Lambda_{TC}$, all the way to the ETC scale,
$\Lambda_{ETC}$. In the limit of extreme walking, one obtains
\begin{equation}
\langle \bar{U}_L U_R \rangle_{\Lambda_{ETC}} =
\left(\frac{\Lambda_{ETC}}{\Lambda_{TC}}\right)^{\gamma_m}
\langle \bar{U}_L U_R \rangle_{\Lambda_{TC}}~.
\label{eqn:enhance}
\end{equation}

We may now use (\ref{eqn:enhance}) to quantify the enhancement of the technicolor condensate required to produce the observed quark masses in a walking model.  Specifically, we will investigate the size of the quark mass which can be achieved in the limit of extreme walking for various $\gamma_m$, and an ETC scale of $10^3$ TeV (which, as shown above, should suffice to meet the CP-conserving
FCNC constraints in the $K$ - and $D$-meson systems).   The calculation requires an estimate of the technicolor scale $\Lambda_{TC}$ and  the technicolor condensate renormalized at the electroweak scale, $\langle \bar{U}_L U_R \rangle_{\Lambda_{TC}}$.

Two estimates of the scales associated with technicolor chiral symmetry breaking are commonly
used in the literature: Naive Dimensional Analysis (NDA) \cite{Weinberg:1978kz,Manohar:1983md,Georgi:1985hf}
and simple dimensional analysis (DA) as applied in \cite{Eichten:1979ah}. In Naive Dimensional Analysis, one
associates $\Lambda_{TC}$ with the "chiral symmetry breaking scale" for the technicolor
theory,  $\Lambda_{TC}=\Lambda_{\chi SB}\approx4\pi v$ and 
$\langle \bar{U}_L U_R \rangle_{\Lambda_{TC}}\approx 4\pi v^3 \approx (580\, {\rm GeV})^3$, (where $v\approx 250$ GeV is the analog of $f_\pi$ in QCD). In
the simple dimensional estimates 
one simply assumes that all technicolor scales are given by $\Lambda_{TC} \approx
1$ TeV, and hence $\langle \bar{U}_L U_R \rangle_{\Lambda_{TC}}\approx (1\, {\rm TeV})^3$.

In Table \ref{tab:two} we estimate the size of
quark mass corresponding to various (constant) values of $\gamma_m$ and an ETC
scale of $10^3$ TeV. We show these values in the range $0 \le \gamma_m \le 2.0$
since $\gamma_m\simeq 0$ in a "running" technicolor theory, and  conformal group representation unitarity implies 
that $\gamma_m \le 2.0$ \cite{Mack:1975je}. 
The usual Schwinger-Dyson analysis used to analyze technicolor theories would
imply that $\gamma_m \le 1.0$ in walking technicolor theories \cite{Holdom:1981rm,Holdom:1984sk,Yamawaki:1985zg,Appelquist:1986an,Appelquist:1986tr,Appelquist:1987fc}, while the values $1.0 \le
\gamma_m \le 2.0$
could occur in strong-ETC theories \cite{Appelquist:1988as,Takeuchi:1989qa,Miransky:1988gk,Fukano:2010yv}.

\begin{table}[h]
\caption{Size of the quark mass $m_q$ generated by technicolor dynamics assuming an ETC scale $\Lambda_{ETC} = 1000$ TeV and various values for the anomalous dimension $\gamma_m$ of the mass operator. In the row labeled NDA [DA], the value of the techniquark condensate at the technicolor scale is taken to be $\langle \bar{T} T \rangle \approx (580\, {\rm GeV})^3$ [$(1000\, {\rm GeV})^3$].  Values of $\gamma_m$ of 1.0 or less correspond to walking theories
\protect\cite{Holdom:1981rm,Holdom:1984sk,Yamawaki:1985zg,Appelquist:1986an,Appelquist:1986tr,Appelquist:1987fc}; values greater than 1.0 correspond to strong-ETC theories \protect\cite{Appelquist:1988as,Takeuchi:1989qa,Miransky:1988gk,Fukano:2010yv}.}
\begin{center}
\tiny
\begin{tabular}{|c||c|c|c|c|c||c|c|c|c|} \hline
$\phantom{\dfrac{\strut}{\strut}} \gamma_m$ &0 & 0.25 & 0.5 & 0.75 & 1.0 & 1.25& 1.5 & 1.75 & 2.0 \\  \hline \hline
$\phantom{\dfrac{\strut}{\strut}} m^{NDA}_q$  \ \ & \ \  0.2 MeV \ \ & \ \ 0.8 MeV \ \ & \ \ 3.5 MeV \ \ & \ \ 15 MeV \ \ & \ \ 63 MeV \ \ & \ \ 260 MeV \ \ & \ \ 1.1 GeV \ \ & \ \ 4.7 GeV \ \ & \ \ 20 GeV\\  \hline 
$\phantom{\dfrac{\strut}{\strut}} m^{DA}_q$  \ \ & \ \  1 MeV \ \ & \ \ 5.6 MeV \ \ & \ \ 32 MeV \ \ & \ \ 180 MeV \ \ & \ \ 1 GeV \ \ & \ \ 5.6 GeV \ \ & \ \ 32 GeV \ \ & \ \ 180 GeV \ \ & \ \ 1 TeV\\  \hline 

\end{tabular}
\end{center}
\label{tab:two}
\end{table}


\section{Discussion}\label{sec:discussion}

Examining Table \ref{tab:two}, we see that generating the charm quark mass from
ETC dynamics at a scale of order $10^3$ TeV requires an anomalous dimension $\gamma_m$
close to or exceeding one, even in the case of the more generous DA estimate of the
technifermion condensate.  It is therefore important for nonpertubative studies of strong technicolor dynamics to determine how large $\gamma_m$ can be in specific candidate theories of walking technicolor. 
Lattice Monte Carlo studies to date \cite{Appelquist:2009ka,Deuzeman:2009mh,Fodor:2009wk,Nagai:2009ip,Svetitsky:2009pz,Bursa:2009we,DelDebbio:2010hx,DeGrand:2009hu,Hasenfratz:2010fi,DeGrand:2010na,delDebbio} prefer values of $\gamma_m \laem 1.0$ in the
theories studied so far. Values of $\gamma_m$ substantially less than one would require a lower ETC scale, which would necessitate the construction of ETC theories with approximate flavor symmetries \cite{Chivukula:1987py,D'Ambrosio:2002ex,Appelquist:2004ai} and corresponding GIM-like partial cancellations of flavor-changing contributions.

If a "walking" theory with $\gamma_m \gaem 1$ is found, then a number of interesting questions should
also be investigated, including:
\begin{itemize}

\item What is the complete phase diagram for theories of this sort, as a function of
the number of "colors" and "flavors" \cite{Banks:1981nn}?

\item Can $\gamma_m$ be larger than one?

\item What is the value of the electroweak $S$ 
\cite{Peskin:1990zt,Peskin:1991sw} parameter\footnote{See \protect\cite{Sannino:2010ca}
for a recent conjecture on this topic.}?

\item Is there a (pseudo-)dilaton with Higgs-like couplings\footnote{For recent discussions
in this regard, see \protect\cite{Dietrich:2005jn,Appelquist:2010gy,Hashimoto:2010nw,Vecchi:2010aj}.}?

\item What are the properties of the lightest vector-mesons which would appear
in $WW$ scattering?

\item Are there other marginal or relevant operators, and can they be useful in
generating quark masses {\'a} la strong-ETC \cite{Appelquist:1988as,Takeuchi:1989qa,Miransky:1988gk,Fukano:2010yv}?

\end{itemize}

Results presented at this conference \cite{delDebbio} are intriguing, and we look forward to
a thorough exploration of the properties of candidate theories of dynamical electroweak
symmetry breaking.

\newpage

\section{Acknowledgements}

This work was supported in part by the US National Science Foundation under grants PHY-0354226
and PHY-0854889.  RSC and EHS gratefully acknowledge conversations with Tom Appelquist, Tom DeGrand, 
Luigi Del Debbio, Francesco Sannino, 
Bob Shrock, Ben Svetitsky, and Rohana Wijewhardana, as well as participants in the
``Strong Coupling Beyond the Standard Model" workshop held at the Aspen Center for Physics
during May and June, 2010. We also thank the Aspen Center for Physics for its support while
this work was completed.

\end{document}